\documentclass[12pt, a4]{article}
\usepackage{parskip}
\usepackage{natbib}
\usepackage[colorlinks=true, urlcolor=blue, linkcolor=blue, citecolor=blue]{hyperref}
\usepackage{amsmath,amsthm,amssymb} 
\usepackage{graphics}
\usepackage{caption}
\usepackage{subcaption}
\usepackage{enumitem}
\usepackage{color}
\usepackage[table]{xcolor}
\usepackage[english]{babel}
\usepackage{xr}
\usepackage[margin=1in]{geometry}
\usepackage{authblk}
\usepackage{secdot}
\usepackage{setspace}
\usepackage{titlesec}
\usepackage{tikz}
\usepackage{bm}
\usepackage{mdframed}
\usepackage{booktabs} 
\usepackage{array}    
\usepackage{comment}
\usepackage{mathtools}
\usepackage{cancel}
\usepackage[calc]{picture}

\usetikzlibrary{arrows.meta,shapes}
\usetikzlibrary{arrows,shapes.arrows,shapes.geometric,shapes.multipart,
decorations.pathmorphing,positioning,shapes.swigs}
\newcolumntype{?}{!{\vrule width 1pt}}

\DeclareMathOperator*{\expit}{expit}
\newcommand{\bc}{\mathbf{c}}
\newcommand{\bC}{\mathbf{C}}

\newcommand{\bL}{\mathbf{L}}

\newcommand{\dP}{\mathrm{dP}}

\titlespacing*{\section}
{0pt}{1\baselineskip}{1\baselineskip}
\titlespacing*{\subsection}
{0pt}{1\baselineskip}{1\baselineskip}

\newcommand\independent{\protect\mathpalette{\protect\independenT}{\perp}}
\def\independenT#1#2{\mathrel{\rlap{$#1#2$}\mkern2mu{#1#2}}}

\titleformat{\section}
{\normalfont\large\bfseries}{\thesection}{1em}{}
\titleformat{\subsection}
{\normalfont\normalsize\bfseries}{\thesubsection}{1em}{}

\theoremstyle{plain}
\newtheorem{definition}{Definition}

\newtheorem{proposition}{Proposition}

\theoremstyle{definition}

\newtheorem{remark}{Remark}

\begin{document}

\title{On treating right-censoring events like treatments }



\author[1]{Lan Wen}
\author[2]{Aaron L. Sarvet}
\author[3,4]{Jessica G. Young}

\affil[1]{Department of Statistics and Actuarial Science, University of Waterloo}
\affil[2]{Department of Biostatistics \& Epidemiology,
University of Massachusetts, Amherst, USA}
\affil[3]{Department of Population Medicine, Harvard Medical School and Harvard Pilgrim Health Care Institute, Boston, Massachusetts}
\affil[4]{Department of Epidemiology, Harvard T.H. Chan School of Pubic Health, Boston, Massachusetts}

\date{}
\maketitle
\begin{abstract}
   \noindent In causal inference literature, potential outcomes are often indexed by the ``elimination of all right-censoring events,'' leading to the perception that such a restriction is necessary for defining well-posed causal estimands. In this paper, we clarify that this restriction is not required: a well-defined estimand can be formulated without indexing on the elimination of such events. Achieving this requires a more precise classification of right-censoring events than has historically been considered, as the nature of these events has direct implications for identification of the target estimand. We provide a framework that distinguishes different types of right-censoring events from a causal perspective, and demonstrate how this framework relates to censoring definitions and assumptions in classical survival analysis literature. By bridging these perspectives, we provide a clearer understanding of how to handle right-censoring events and provide guidance for identifying causal estimands when right-censored events are present.
\end{abstract}

\section{Introduction}
Right-censoring events are ubiquitous in randomized trials and observational studies, and they pose significant challenges for estimating the average causal effect of an exposure or treatment on an outcome of interest. The means in which right-censoring events are addressed or handled has important implications for both study design \citep{guideline2017addendum} and the validity of subsequent statistical inferences.

In the causal inference literature, a frequent approach to handling right-censoring is to conceptualize an intervention that removes all right-censoring events and by indexing potential outcomes under that hypothetical elimination \citep{neugebauer2014targeted,richardson2014causal,young2020causal,hernan2020whatif,wen2021parametric,wen2022multiply,wen2023intervention}. 
While useful for certain identification arguments, this framing can be misconstrued as asserting that removal of all right-censoring events is necessary to define a well-posed or well-defined causal estimand.

By contrast, classical survival analysis typically does not invoke hypothetical interventions that eliminate right-censoring events. Instead, the standard approach assumes conditionally independent right-censoring \citep{Fleming,kalbfleisch2002statistical}, an assumption that has likewise been adopted in several causal inference settings \citep{cai2020one,rytgaard2021estimation,ying2022proximal,cui2023estimating,li2025regression,bu2025semiparametric}. 
The differences of approaches (treating right-censoring event as preventable in causal framework, versus assuming conditional independence in classical survival analysis) reflects a conceptual gap that can lead to confusion about what estimands are being defined and when they are identifiable.

In this paper we close that gap. We show that a well-defined causal estimand need not be indexed on the elimination of right-censoring events: one can formulate estimands that do not require positing a joint intervention to remove all of these events. Doing so, however, requires a more careful classification of right-censoring events. 
Different kinds of right-censoring events (e.g., administrative end-of-study censoring, loss to follow-up due to relocation, or right-censoring that is a direct consequence of treatment) can have distinct causal and identification implications. 
Consequently, handling these events should be driven by the (causal) estimand of interest: the estimand determines which right-censoring events require elimination and, in turn, which identification and estimation methods are appropriate.

Our contributions are threefold. First, we introduce a framework that distinguishes types of right-censoring events and characterizes their roles relative to the target estimand. 
Second, we show when and under what assumptions various estimands are identifiable, so that right-censoring events are handled transparently and appropriately for the scientific question at hand.
Third, we relate this framework to classical survival-analysis assumptions. In doing so, we show that classical formulations of independent censoring do not explicitly engage with causal models, while modern causal inference approaches often imply that estimands should be defined through interventions on right-censoring events, even when right-censoring events has no causal effect on the outcome. 
Neither is one-size-fits-all; rather, each provides tools appropriate for different estimands and right-censoring mechanisms.

The remainder of the paper is organized as follows. Section 2 and 3 introduces notation and estimands considered herein. Section 4 gives formal definitions of different right-censoring event types within the causal framework. Section 5 states the identification results under various estimands and discusses practical solutions to violations of assumptions. Section 6 connects these results to classical survival assumptions, and Section 7 illustrates the ideas with examples via simulation studies. 

\section{Longitudinal data structure}
\sloppy\allowdisplaybreaks

Consider a longitudinal study in which measurements on a random sample of individuals from a population of interest are collected over a follow-up period.  Let $k=0,\ldots,K$ denote equally spaced intervals within this follow-up, where $k=0$ the baseline interval and $K$ the last interval in which individual measurements are available. The length of these intervals will depend on the study design.  For instance, in a study using electronic health records with dates of visits, services, labs, or prescriptions, $k$ might denote a single day of the follow-up.  By contrast, in a trial or cohort study with scheduled follow-up visits no more frequent than weekly, $k$ would represent a week. 

Following \cite{robins2010alternative}, we define \textit{factual (or realized) variables} as those that could, in principle, be recorded in the actual world for subjects participating in the study.
In turn, let $\bL_k$ denote a vector of time-varying covariates relative to time interval $k$, $A_k$ the \textsl{treatment of substantive interest} to the investigator in time interval $k$, and $Y_k$ an indicator of survival by time interval $k$ ($Y_k=1$ if the individual remains alive at time $k$ and $Y_k=0$ otherwise), for $k\geq 0$.  We will use the convention of an overline and underline to denote history and future of a random variable, respectively; e.g., $\overline{Y}_k=(Y_0,Y_1,\ldots,Y_k)$ and $\underline{Y}_{k+1}=(Y_{k+1},Y_{k+1},\ldots,Y_K)$. 
Correspondingly, let $C_k$ denote an indicator of an event by time $k$ ($k=1,\ldots,K$) that terminates the observation process of realized variables such that all information on the participant is unknown to the investigator (i.e. missing) beyond time $k$.
 In general, there can be multiple of these events such that $\mathbf{C}_k = (C_{k1},C_{k2},\ldots)$, for $k=1,\ldots,K,$ each of which may affect the outcome differently or not at all. We refer to such events collectively as ``\textit{observation-terminating events}''. 

Examples of observation-terminating events include deletion of a study app from a participant's electronic device \citep{adu2020user,amagai2022challenges}, or a formal withdrawal from a trial \citep{gabriel2011data,sofrygin2019targeted,wilson2021quantifying}, after which no further data are recorded for that individual.
Events arising from designs analogous to Type I and Type II stopping rules in survival analysis \citep{Fleming,kalbfleisch2002statistical,andersen2012statistical,klein2014handbook} likewise constitute observation-terminating events.
Throughout, we will rely on the assumption that either the temporal ordering $(\bC_k,Y_k,\bL_k,A_k)$ holds in each interval $k$ or that the length of each interval $k$ is selected to be so small that there is no dependence between events within an interval (akin to the ``no ties'' assumption in survival analysis). We define $\bC_0\equiv Y_0\equiv 0$; that is, we restrict the population to those who are alive and and have not yet experienced an event that precludes observation of future survival processes. 

By the above, we can summarize the observed data for any individual who remains alive and under observation through the end of the study period (i.e., $\overline{\bC}_K=0,\overline{Y}_K=1)$ as a vector of factual variables given by $(\overline{\bL}_{K-1},\overline{A}_{K-1},\overline{\bC}_K,\overline{Y}_K)$. 
For an individual who died in an interval $k<K$ without any observation-terminating event occurring prior to that time (i.e., $\overline{\bC}_k=0,\overline{Y}_{k-1}=1,\underline{Y}_k=0)$, we observe $(\overline{\bL}_{k-1},\bar{A}_{k-1}\overline{\bC}_K,\overline{Y}_{K})$.  Otherwise, for an individual whose observation was terminated before prior to $k$ while still alive (i.e., $\overline{Y}_{k-1}=1, \overline{\bC}_{k-1}=0,C_k=1$), we observe $(\overline{\bL}_{k-1},\bar{A}_{k-1}\overline{\bC}_K,\overline{Y}_{k-1})$.  



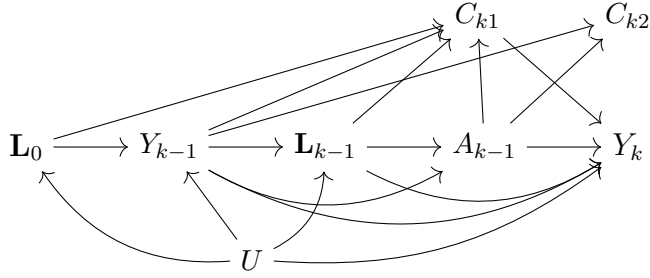
\begin{figure}[htbp]
\begin{minipage}{1\linewidth}
\centering
 \begin{tikzpicture}
\tikzset{
  line width=1pt, outer sep=0pt,
  ell/.style={draw,fill=white,inner sep=1pt,line width=1pt},
  swig vsplit={gap=8pt,inner line width right=0.15pt}
}
\begin{scope}[every node/.style={thick,draw=none}]
\node (L0) {$\mathbf L_0$};

\node[right=of L0] (Ykm1) {\small $Y_{k-1}$};
\node[right=of Ykm1] (Lkm1) {\small $\mathbf L_{k-1}$};
\node[right=of Lkm1] (Akm1) {\small $A_{k-1}$};
\node[right=of Akm1] (Yk) {$Y_k$};

\node (rCk1) at (6,1.75) {\small $C_{k1}$};
\node (rCk2) at (8,1.75) {\small $C_{k2}$};
\node (U) at (3,-1.5) {\small $U$};

\draw[->] (Ykm1) -- (Lkm1); 
\draw[->] (Ykm1) edge[bend right=30] (Yk); 

\draw[->] (Ykm1) edge[bend right=30] (Akm1); 
\draw[->] (Lkm1) edge[bend right=30] (Yk);      


\draw[->] (L0) -- (Ykm1);  
\draw[->] (L0) -- (rCk1);  
\draw[->] (Lkm1) -- (rCk1); 
\draw[->] (Lkm1) -- (Akm1); 
\draw[->] (Akm1) -- (rCk1); 
\draw[->] (Ykm1) -- (rCk1); 
\draw[->] (rCk1) -- (Yk); 

\draw[->] (Akm1) -- (rCk2); 
\draw[->] (Ykm1) -- (rCk2); 

\draw[->] (Akm1) -- (Yk); 

\draw[->] (U)  edge[bend left=30] (L0); 
\draw[->] (U) -- (Ykm1); 
\draw[->] (U) edge[bend right=30] (Lkm1); 
\draw[->] (U) edge[bend right=20] (Yk); 
\end{scope}
\end{tikzpicture}
\caption{\label{fig:figure00}Directed Acyclic Graph (DAG) depicting the factual data. For simplicity, we define two observation-terminating event $\bC_k = (C_{k1},C_{k2})$. Note that we omit other arrows (e.g., $\mathbf L_0$ to $Y_{k-1}$) to reduce clutter, as adding omitted edges from past to future variables does not affect our assumptions.}
\end{minipage}
\end{figure}

\section{Two distinct estimands}
\label{sec:estimands}
Suppose the investigators who conducted the study described above communicated their goal as interest in the ``cumulative survival probability by some $k$ under treatment regime $g$'', where $g = (g_0,\ldots,g_{K-1})$ denotes a deterministic sequence of interventions on $A_k$ (see \citealp{Young2011,young2019inverse,wen2023intervention} for a formal definition).  More generally, we can formally define an average causal effect as a contrast in average potential outcomes between two regimes $g=g'$ and $g=g{''}$.  
For ease of exposition, we consider the average causal effect given by a contrast between the cumulative probabilities of survival under two treatment regimes $g' = (g_0',\ldots,g_{K-1}')$ and $g'' = (g_0'',\ldots,g_{K-1}'')$, where $g_k' = a_k'$ and $g_k'' = a_k''$ for each $k = 0,\ldots,K-1$, with $\bar{a}_K'\neq \bar{a}_K''$. These results extend to any treatment strategies $g$ that may depend on past covariates.

For any $k=1,\ldots,K$, first consider the potential outcome $Y_k^{g,\overline{\bc}=0}$, the survival status an individual would have at time $k$ had that individual followed treatment regime $g$ and, further, had (somehow) all observation-terminating events been eliminated through $k$. This potential outcome can be more explicitly defined relative to a causal model. Figure \ref{fig:figure00} is a causal directed acyclic graph (DAG) depicting a possible underlying dependency structure among the factual variables in this study. Throughout we will assume that statistical independencies in the data are faithful to the graph under study \citep{verma2022equivalence}.

Based on the exposition above, the Single World Intervention Graph (SWIG; \citealp{richardson2013single}) in Figure \ref{fig:figureswig1} is a node-splitting transformation of the causal DAG in Figure \ref{fig:figure00}, explicitly depicting $(g,\overline{\bc})$ potential variables.  In turn, one possible formalization of the ``cumulative survival probability by some $k$ under treatment regime $g$'' is given by:
\begin{equation}
\Psi_1^g \coloneqq E(Y_k^{g,\bar{\bc}=0})\label{tointervene}
\end{equation}
 Alternatively, consider the potential outcome $Y_k^{g}$, the survival status an individual would have at time $k$ had that individual followed treatment regime $g$ but with no additional interventions to eliminate observation-terminating events $\bC_k$. This distinct potential outcome is more explicitly represented by the SWIG in Figure \ref{fig:figureswig2}, which clarifies that $Y_k^{g}$ corresponds to then individual's survival status at $k$ had they followed treatment regime $g$ \textit{and} had they experienced their \textsl{natural} observation termination process under $g$ \citep{richardson2013single,young2020causal}. Correspondingly, an alternative formalization of the ``cumulative survival probability by $k$ under treatment regime $g$'' is a given by: 
\begin{equation}
\Psi_2^g \coloneqq E(Y_k^{g})\label{nottointervene}
\end{equation}


The estimands \eqref{tointervene}  and \eqref{nottointervene} are generally not equal. As shown in Section \ref{sec:estimandident}, they coincide in two special cases: (i) when no individual's observation process is terminated before time $K$; or (ii) when the observation-terminating events in $\bC_k$ have no direct causal effect on $\underline{Y}_k$. The latter corresponds to the setting illustrated in Figure \ref{fig:figure00} if, like $C_{k2}$, $C_{k1}$ has no arrow directed into $Y_k$.

\begin{figure}[htbp]
\begin{minipage}{1\linewidth}
\centering
 \begin{tikzpicture}
\tikzset{
  line width=1pt, outer sep=0pt,
  ell/.style={draw,fill=white,inner sep=1pt,line width=1pt},
  swig vsplit={gap=8pt,inner line width right=0.15pt}, node distance=1.25cm
}
\begin{scope}[every node/.style={thick,draw=none}]
\node (L0) {$\mathbf L_0$};

\node[right=of L0] (Ykm1) {\small $Y_{k-1}^{g,\bar{\bc}=0}$};
\node[right=of Ykm1] (Lkm1) {\small $\mathbf L_{k-1}^{g,\bar{\bc}=0}$};
\node[right=of Lkm1] (Akm1)  {$A_{k-1}^{g,\bar{\bc}=0}$};
        \node (midA1)[right = -0.2 of Akm1,    anchor=west] {$\mid$};
        \node (a)  [right = -0.2 of midA1, anchor=west] {$a_{k-1}$}; 

\node[right=of a] (Yk) {$Y_k^{g,\bar{\bc}=0}$};

\node (rCk1) at (7.5,1.75) {$C_{k1}^{g,\bar{\bc}=0}$};
        \node (midC1)[right =  -0.2 of rCk1,    anchor=west] {$\mid$};
        \node (ck1)  [right =  -0.2 of midC1, anchor=west] {\small $c_{k1}=0$}; 

\node (rCk2) at (10.5,1.75) {$C_{k2}^{g,\bar{\bc}=0}$};
        \node (midC2)[right =  -0.2 of rCk2,    anchor=west] {$\mid$};
        \node (ck2)  [right =  -0.2 of midC2, anchor=west] {\small $c_{k2}=0$}; 

\node (U) at (3,-1.5) {\small $U$};

\draw[->] (Ykm1) -- (Lkm1); 
\draw[->] (Ykm1) edge[bend right=30] (Yk); 

\draw[->] (Ykm1) edge[bend right=30] (Akm1); 
\draw[->] (Lkm1) edge[bend right=30] (Yk);      


\draw[->] (L0) -- (Ykm1);  
\draw[->] (L0) -- (rCk1);  
\draw[->] (Lkm1) -- (rCk1); 
\draw[->] (Lkm1) -- (Akm1); 
\draw[->] (a) -- (rCk1); 
\draw[->] (Ykm1) -- (rCk1); 
\draw[->] (ck1) -- (Yk); 

\draw[->] (a) -- (rCk2); 
\draw[->] (Ykm1) -- (rCk2); 

\draw[->] (a) -- (Yk); 

\draw[->] (U)  edge[bend left=30] (L0); 
\draw[->] (U) -- (Ykm1); 
\draw[->] (U) edge[bend right=30] (Lkm1); 
\draw[->] (U) edge[bend right=20] (Yk); 
\end{scope}
\end{tikzpicture}
\caption{\label{fig:figureswig1}Single World Intervention Graph (SWIG) depicting ($g,\overline{\bc})$ potential variables (counterfactuals). For simplicity, we define two observation-terminating event $\bC_k = (C_{k1},C_{k2})$. Note that we omit other arrows (e.g., $\mathbf L_0$ to $Y_{k-1}$) to reduce clutter, as adding omitted edges from past to future variables does not affect our assumptions.}
\end{minipage}
\end{figure}

\begin{figure}[htbp]
\begin{minipage}{1\linewidth}
\centering
 \begin{tikzpicture}
\tikzset{
  line width=1pt, outer sep=0pt,
  ell/.style={draw,fill=white,inner sep=1pt,line width=1pt},
  swig vsplit={gap=8pt,inner line width right=0.15pt},
  node distance=1.5cm
}
\begin{scope}[every node/.style={thick,draw=none}]
\node (L0) {$\mathbf L_0$};

\node[right=of L0] (Ykm1) {\small $Y_{k-1}^g$};
\node[right=of Ykm1] (Lkm1) {\small $\mathbf L_{k-1}^g$};
\node[right=of Lkm1] (Akm1)  {$A_{k-1}^g$};
        \node (midA1)[right = -0.2 of Akm1,    anchor=west] {$\mid$};
        \node (a)  [right = -0.2 of midA1, anchor=west] {$a_{k-1}$}; 

\node[right=of a] (Yk) {$Y_k^g$};

\node (rCk1) at (8,1.75) {$C_{k1}^g$};
\node (rCk2) at (10,1.75) {$C_{k2}^g$};

\node (U) at (3,-1.5) {\small $U$};

\draw[->] (Ykm1) -- (Lkm1); 
\draw[->] (Ykm1) edge[bend right=30] (Yk); 

\draw[->] (Ykm1) edge[bend right=30] (Akm1); 
\draw[->] (Lkm1) edge[bend right=30] (Yk);      


\draw[->] (L0) -- (Ykm1);  
\draw[->] (L0) -- (rCk1);  
\draw[->] (Lkm1) -- (rCk1); 
\draw[->] (Lkm1) -- (Akm1); 
\draw[->] (a) -- (rCk1); 
\draw[->] (Ykm1) -- (rCk1); 
\draw[->] (rCk1) -- (Yk); 

\draw[->] (a) -- (rCk2); 
\draw[->] (Ykm1) -- (rCk2); 

\draw[->] (a) -- (Yk); 

\draw[->] (U)  edge[bend left=30] (L0); 
\draw[->] (U) -- (Ykm1); 
\draw[->] (U) edge[bend right=30] (Lkm1); 
\draw[->] (U) edge[bend right=20] (Yk); 
\end{scope}
\end{tikzpicture}
\caption{\label{fig:figureswig2}Single World Intervention Graph (SWIG) depicting $g$ potential variables (counterfactuals). For simplicity, we define two observation-terminating event $\bC_k = (C_{k1},C_{k2})$. Note that we omit other arrows (e.g., $\mathbf L_0$ to $Y_{k-1}$) to reduce clutter, as adding omitted edges from past to future variables does not affect our assumptions.}
\end{minipage}
\end{figure}
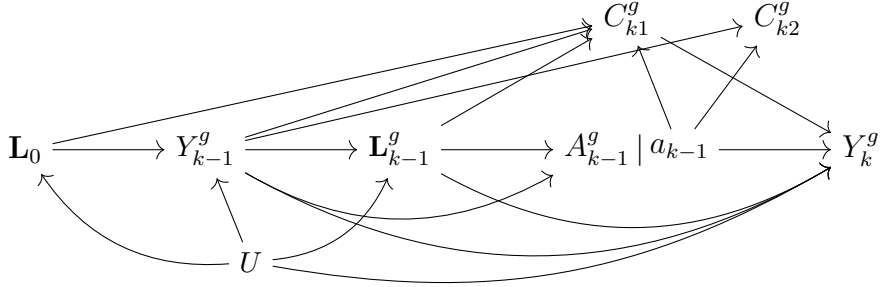

\section{{Defining and classifying right-censoring events}}
\label{sec:comprehensivedef}

To classify the observation-terminating events in $\bC_k$ ($k = 1, \ldots, K$) introduced in the previous section, we draw on the definition of a ``right-censoring event" introduced in \cite{young2020causal}:
\begin{quote}
    ``A \textit{[right-]censoring event} is any event occurring in the study by [time interval] $k$, $k=1,\ldots,K$, that ensures the values of \textit{all future potential outcomes of interest} under [a point treatment level] $a$ are unknown even for an individual receiving the intervention $a$.''
\end{quote}
Referring back to the estimands in Section \ref{sec:estimands}, as noted by \cite{young2020causal}, events in $\bC_k$ meets this definition of a right-censoring event: whenever any event in $\bC_k$ occurs by time $k$, it prevents us from observing all subsequent potential outcomes under the point treatment level $a$ for that individual.
As discussed later, how these events are addressed depends on the estimand of interest.
Building on the work of \cite{young2020causal}, we propose a general definition of right-censoring that applies uniformly to any outcome of interest and that facilitates the formulation of clear identification assumptions.

\subsection{A unified definition of right-censoring event}
The definition proposed by \citet{young2020causal} is inherently tied to the outcomes of interest and requires that the causal estimand be clearly specified prior to data analysis. As a result, the choice of estimand automatically identifies which events are treated as part of the censoring mechanism and how they should be handled.
Adopting this perspective, we present a more general definition of right-censoring event that is not limited to potential outcomes of interest under $a$:

\begin{definition}[Generalized definition of right-censoring event]
For any subject in a study, a right-censoring event refers to any event occurring in the study by time interval $k$, for $k = 1, \ldots, K$, that renders the values of their \textit{future outcomes of interest} unobservable.
\end{definition}
To highlight the differences between $\Psi_1^g$ and $\Psi_2^g$, we further classify different right-censoring events into \textit{nuisance right-censoring events}, and all other right-censoring events as \textit{non-nuisance right-censoring events}. 

\begin{definition}[Nuisance and non-nuisance right-censoring events]
Any right-censoring event in which the investigator's causal question or ideal trial does \textit{not} invoke the \textit{elimination} or \textit{prevention} of the event is referred to as a ``{nuisance right-censoring event}.'' All other censoring mechanisms—those that are integral to the definition of the estimand—will be considered ``{non-nuisance right-censoring events}''.
\end{definition}

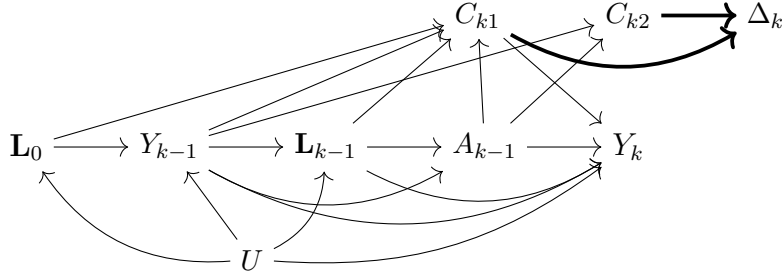
\begin{figure}[htbp]
\begin{minipage}{1\linewidth}
\centering
 \begin{tikzpicture}
\tikzset{
  line width=1pt, outer sep=0pt,
  ell/.style={draw,fill=white,inner sep=1pt,line width=1pt},
  swig vsplit={gap=8pt,inner line width right=0.15pt}
}
\begin{scope}[every node/.style={thick,draw=none}]
\node (L0) {$\mathbf L_0$};

\node[right=of L0] (Ykm1) {\small $Y_{k-1}$};
\node[right=of Ykm1] (Lkm1) {\small $\mathbf L_{k-1}$};
\node[right=of Lkm1] (Akm1) {\small $A_{k-1}$};
\node[right=of Akm1] (Yk) {$Y_k$};

\node (rCk1) at (6,1.75) {\small $C_{k1}$};
\node (rCk2) at (8,1.75) {\small $C_{k2}$};
\node [right=of rCk2] (Cktilde) {\small $\Delta_{k}$};

\node (U) at (3,-1.5) {\small $U$};

\draw[->] (Ykm1) -- (Lkm1); 
\draw[->] (Ykm1) edge[bend right=30] (Yk); 

\draw[->] (Ykm1) edge[bend right=30] (Akm1); 
\draw[->] (Lkm1) edge[bend right=30] (Yk);      


\draw[->] (L0) -- (Ykm1);  
\draw[->] (L0) -- (rCk1);  
\draw[->] (Lkm1) -- (rCk1); 
\draw[->] (Lkm1) -- (Akm1); 
\draw[->] (Akm1) -- (rCk1); 
\draw[->] (Ykm1) -- (rCk1); 
\draw[->] (rCk1) -- (Yk); 

\draw[->] (Akm1) -- (rCk2); 
\draw[->] (Ykm1) -- (rCk2); 

\draw[->] (Akm1) -- (Yk); 

\draw[->] (U)  edge[bend left=30] (L0); 
\draw[->] (U) -- (Ykm1); 
\draw[->] (U) edge[bend right=30] (Lkm1); 
\draw[->] (U) edge[bend right=20] (Yk); 

\draw[->] (rCk1) edge[line width=0.5mm, bend right=30] (Cktilde); 
\draw[->] (rCk2) edge[line width=0.5mm] (Cktilde); 

\end{scope}
\end{tikzpicture}
\caption{\label{fig:figure0}Directed Acyclic Graph (DAG) depicting the factual data, with right-censoring state added to Figure \ref{fig:figure00}. Bold arrows are used to emphasize deterministic relationships assumed herein.}
\end{minipage}
\end{figure}

An example of non-nuisance right-censoring event is non-adherence to $g$, when the outcome of interest is defined under a regime where non-adherence is entirely prevented. In contrast, nuisance right-censoring events include those such as loss to follow-up or study withdrawal, when those events are not part of the intervention requiring their elimination. 
In light of our definitions, let $\Delta_k$ denote the \textit{non-observation state} ($\Delta_k$=1 if $\underline{Y}_k$ is unobservable and 0 otherwise) in time interval $k$, $k=1,\ldots,K$. Specifically, $$\Delta_k = \max (\bC_k).$$
Figure \ref{fig:figure0} extends the DAG in Figure \ref{fig:figure00} to include $\Delta_k$.  
Non-nuisance right-censoring events should be defined in terms of the substantive causal question/story and the specific hypothetical interventions that would prevent them; only with such specification does it make sense to invoke a consistency assumption that links potential outcomes under intervention on those events to the factual data \citep{young2024story}.
In contrast, nuisance right-censoring events are not targeted by the estimand, and identification proceeds without invoking such a consistency assumption for each nuisance event.

\begin{figure}[htbp]
\centering
\begin{tikzpicture}[font=\small]
\begin{scope}[shift={(-3,0)}] 
  \fill[gray!10] (0,0) rectangle (4,3);
  \draw[thick, rounded corners=4pt] (0,0) rectangle (4,3);
  \node[anchor=north] at (1.8,3.1) {{\footnotesize Right-censoring events}};

  \draw[line width=1.5pt, blue!80!black, dashed, rounded corners=5pt] (0.8,0.6) rectangle (3.2,2.4);
  \node at (1.1,2.15) {$\mathbf{C}_k$};
  \node[anchor=north] at (2,-0.2) {(a) $\Psi_1^g\coloneqq E(Y_k^{g,\bar{\bc}=0})$};
\end{scope}

\begin{scope}[shift={(3,0)}] 
  \fill[gray!10] (0,0) rectangle (4,3);
  \draw[thick, rounded corners=4pt] (0,0) rectangle (4,3);
  \node[anchor=north] at (1.8,3.1) {{\footnotesize Right-censoring events}};

  \fill[gray!10, rounded corners=3pt] (0.8,0.6) rectangle (3.2,2.4); 
  \draw[line width=1.5pt, blue!80!black, dashed, rounded corners=5pt] (0.8,0.6) rectangle (3.2,2.4);
  \fill[red!50, opacity=0.5, rounded corners=3pt] (0.8,0.6) rectangle (3.2,2.4);
  \draw[thick, red!80!black, rounded corners=3pt] (0.8,0.6) rectangle (3.2,2.4);
  \node at (1.1,2.15) {$\mathbf{C}_k$};

  \node[anchor=north] at (2,-0.2) {(b) $\Psi_2^g\coloneqq E(Y_k^{g})$};
\end{scope}

\begin{scope}[shift={(0,-1)}] 
  \filldraw[red!50, opacity=0.5, draw=black, rounded corners=2pt] (-2,-0.2) rectangle (-1.6,-0.6);
  \node[right, anchor=west] at (-1.5,-0.5) {Nuisance};
  \filldraw[gray!10, draw=black, rounded corners=2pt] (0.6,-0.2) rectangle (1,-0.6);
  \node[right, anchor=west] at (1.1,-0.5) {Non-nuisance};
  \draw[line width=1.5pt, blue!80!black, dashed, rounded corners=2pt] (4,-0.2) rectangle (4.4,-0.6);
  \node[right, anchor=west] at (4.5,-0.5) {$\mathbf{C}_k$};
\end{scope}
\end{tikzpicture}
\caption{Classification of right-censoring events relative to the estimands considered herein.}
\label{fig:two-censoring-diagrams}
\end{figure}
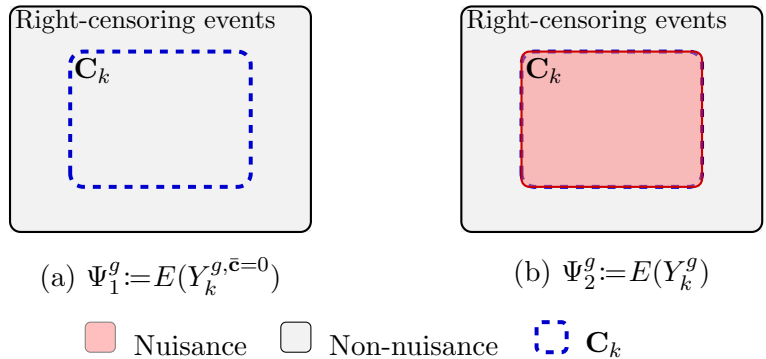

Since events in $\bC_k$ are right-censoring events, each component of $\bC_k$ is classified as belonging either to the set of nuisance right-censoring events or to the set of non-nuisance right-censoring events, relative the the estimand of interest (see Figure \ref{fig:two-censoring-diagrams} and the next section for a detailed discussion).

\begin{proposition}
    All nuisance right-censoring events are a subset of the set of observation-terminating events $\bC_k$.
    \label{prop:nuisance}
\end{proposition}

A proof can be found in Appendix B. Proposition \ref{prop:nuisance} states that if there is a nuisance right-censoring event, then this event has to be a observation-terminating event (i.e., part of $\bC_k$). Moreover, if all observation-terminating events in the study are nuisance right-censoring events, it is sufficient to specify assumptions about the {non-observation state $\Delta_k$} itself -- such as conditional independent $\Delta_k$ (see Sections \ref{sec:estimandident} and \ref{sec:connection}) -- to achieve identification.

\subsection{Relating examples from Section \ref{sec:estimands} to the right-censoring event definition}
\label{sec:illustexamples}

Returning to the estimands in Section \ref{sec:estimands}, suppose for concreteness that the study involves two disjoint observation-terminating events, so that $\bC_k = (C_{k1}, C_{k2})$.
Based on this setup, consider the two estimands we defined in Section \ref{sec:estimands}:
\begin{itemize}[leftmargin=1.2in]
    \item[\textit{Example 1.}] $\Psi_1^g \coloneqq E(Y_k^{g,\bar{\bc}=0})$;
    \item[\textit{Example 2.}] $\Psi_2^g \coloneqq E(Y_k^{g})$.
\end{itemize}
In the next section we study identification of $\Psi_1^g$ in the presence of only non-nuisance right-censoring events. The subsequent section considers identification of $\Psi_2^g$ in the complementary setting with nuisance right-censoring events. 

\section{Identification of causal estimands}
\label{sec:estimandident}
\subsection{Example 1. Identification in the presence of only non-nuisance right-censoring events}
In Example 1, the outcome of interest is a potential variable defined under a joint intervention ($g,\bar{\bc}=0)$. More precisely, $\Psi_1^g$ is the cumulative survival probability at time $k$ if every individual followed treatment regime $g$ \textit{and} all observation-terminating events ($C_{k1}$ and $C_{k2}$) were eliminated. In this example the non-nuisance right-censoring events at time $k$ include $I(A_{k-1}\neq g_{k-1})$, $C_{k1}$ and $C_{k2}$; there are no nuisance right-censoring events. For this estimand, $I(A_{k-1}\neq g_{k-1})$ is a non-nuisance right-censoring event at time $k$ because, for any individual whose treatment history at time $k$ does not equal $\bar{g}_{k-1}$, their future potential outcomes $\underline{Y}_{k}^{g}$ are unobservable and therefore $\underline{Y}_{k}^{g,\bar{\bc}=0}$ are unobservable. Similar reasoning can be made about $C_{k1}$ and $C_{k2}$.

At this stage it is helpful to describe a hypothetical trial whose unadjusted outcome contrast can be interpreted as the causal estimand of interest. Doing so not only clarifies the objective and makes the estimand explicit \citep{young2024story,moreno2024ideal}, but also helps differentiate it from other possible estimands.
Investigators aiming to identify $\Psi_1^g$ would describe a trial in which all participants are perfectly randomized to one of two treatment regimes, $g'$ and $g''$. In each arm every participant adheres fully to their assigned regime until the end of follow-up, and any observation-terminating events ($C_{k1}$ and $C_{k2}$) are prevented. In other words, all non-nuisance right-censoring events (as defined above) are eliminated. We now turn to the formal identification of $\Psi_1^g$.

\subsubsection{Identifying assumptions for $\Psi_1^g$}

Consider the DAG in Figure \ref{fig:figure00} and the corresponding SWIG in Figure \ref{fig:figureswig1}, which represent the intervention defined in Example 1.
For $k=1,\ldots, K$, we consider the following set of assumptions that are sufficient for identification of $\Psi_1^g \coloneqq E(Y_k^{g,\bar{\bc}=0})$:
\begin{itemize}[label={},leftmargin=0mm]
    \item \textit{Assumption 1.1 }(Conditional $\bC_k$-exchangeability): 
\begin{equation}\underline{Y}_k^{g,\bar{\bc}=0}\independent \bC_{k} \mid (\overline{\mathbf{L}}_{k-1},\overline{A}_{k-1}=\overline{g}_{k-1}, \overline{Y}_{k-1}=1, \overline{\bC}_{k-1}=0);\label{eq:cenoutE}\end{equation} 
    \item \textit{Assumption 1.2} (Conditional treatment exchangeability): $$\underline{Y}_k^{g,\bar{\bc}=0}\independent A_{k} \mid (\overline{\mathbf{L}}_{k-1},\overline{A}_{k-1}=\overline{g}_{k-1}, \overline{Y}_{k-1}=1, \overline{\bC}_{k}=0);$$
    \item \textit{Assumption 1.3} (Consistency): $(\overline\bL_k^{g,\bar{\bc}=0},\overline{Y}_k^{g,\bar{\bc}=0})=(\overline\bL_k,\overline Y_k) ~\text{if}~\bar{A}_{k-1}=\overline{g}_{k-1} ~\text{and} ~\overline{\mathbf{C}}_{k}=0$.
    \item \textit{Assumption 1.4} (Positivity): $P(\overline{A}_{k-2}=\overline{g}_{k-2},\bar{\mathbf{L}}_{k-1}=\bar{\mathbf{l}}_{k-1},\overline{Y}_{k-1}=1,\overline{\Delta}_{k-1}=0)>0\longrightarrow$ $$P(A_{k-1}=g_{k-1},\Delta_k=0\mid \overline{Y}_{k-1}=1,\overline{\Delta}_{k-1}=0,\overline{A}_{k-2}=\overline{g}_{k-2},\bar{\mathbf{L}}_{k-1}=\bar{\mathbf{l}}_{k-1})>0.$$
\end{itemize}

\begin{definition}[Model $\mathcal{M}_1$]
    Let $\mathcal{M}_1$ denote a model that satisfy Assumptions 1.1--1.4.
\end{definition}

\begin{proposition}
 Under models $\mathcal{M}_1$, $\Psi_1^g$ is identified via equation \eqref{eq:ident}:
 \begin{align}
    \int_{\bar{\mathbf{l}}_{k-1}}& \!\!\!\!\!\!\!\!\!\!\!\!\!\!\! E(Y_k\mid \overline{Y}_{k-1}=1,\overline{A}_{k-1}=\overline{g}_{k-1},\bar\Delta_k=0,\overline{\bL}_{k-1}=\bar{\mathbf {l}}_{k-1}) \Big[\prod_{j=1}^{k-1} \dP(\mathbf l_{j}\mid \bar{\mathbf {l}}_{j-1},\overline{Y}_{j}=1,\overline{A}_{j-1}=\overline{g}_{j-1},{\bar{\Delta}}_{j}=0)\nonumber
    \\& P(Y_{j}=1\mid \overline{Y}_{j-1}=1,\overline{A}_{j-1}=\overline{g}_{j-1},\bar\Delta_{j}=0, \overline{\bL}_{j-1}=\bar{\mathbf {l}}_{j-1})\Big] \dP(\mathbf l_0).
    \label{eq:ident}
\end{align}
 \end{proposition}

In the example above, the variables $C_{k1}$ and $C_{k2}$ should be defined with respect to the substantive causal question that would necessarily involve joint interventions on $C_{k1}$ and $C_{k2}$. This substantive question would generally differ from the investigator's original aim described at the start of Section \ref{sec:estimands}.
Although our derivation does not require direct observation of the values of $C_{k1}$ and $C_{k2}$, it relies on knowing the nature of the censoring mechanisms to properly adjust for their shared causes with $Y_k$; the same consideration applies to the next estimand.

\subsection{Example 2. Identification in the presence of nuisance right-censoring events}
\label{sec:example2}

In Example 2, the outcome of interest is a potential variable defined under intervention $g$. More precisely, $\Psi_2^g$ is the cumulative survival probability at time $k$ if every individual followed treatment regime $g$, and corresponds to the substantive target specified by the investigator at the beginning of Section \ref{sec:estimands}. 
Implicit in this interpretation is that observation-terminating events $\bC_k$ are not intervened upon. 
In this setting, the non-nuisance right-censoring event at time $k$ is $I(A_{k-1}\neq g_{k-1})$; nuisance right-censoring events are $C_{k1}$ and $C_{k2}$. 

Investigators aiming to identify $\Psi_2^g$ would describe a hypothetical trial in which all participants are perfectly randomized to one of two treatment regimes, $g'$ and $g''$, and in each arm every participant adheres fully to their assigned regime throughout follow-up.
In this trial, all outcomes are eventually observed, regardless of whether participants discontinue treatment, relocate, or otherwise become unreachable through routine follow-up. For individuals for whom an observation-terminating event in $\bC_k$ occurs, outcomes that are unobserved during the study are nonetheless recoverable post hoc (e.g., via linkage to national death indices; comprehensive digital trace and physical record mining, if resources are unlimited) even though events in $\bC_k$ are not eliminated. 
Implicitly, this corresponds to an intervention on the {non-observation state} $\Delta_k$ which does not directly affect $Y_k$ (i.e, $Y_k = Y_k^{\bar\delta=0}=Y_k^{\bar\delta=0,\bar{\bC}}$; see Figure C.1 in Appendix C).  As a result, the outcome becomes observable through mechanisms that are independent of any observation-terminating events.

\subsubsection{Identifying assumptions for $\Psi_2^g$}
\label{sec:identpsi2}

Now consider the same DAG in Figure \ref{fig:figure00} together with the SWIG in Figure \ref{fig:figureswig2}, corresponding to the intervention in Example 2.

Suppose that, in Figure \ref{fig:figure00}, there are no arrows from $(C_{k1},C_{k2})$ into $Y_k$ at any time point $k$.
Given this structure, we posit the following assumptions for each $k=1,\ldots, K$, which are sufficient for identification of $\Psi_2^g \coloneqq E(Y_k^{g})$:
\begin{itemize}[label={},leftmargin=0mm]
    \item \textit{Assumption 2.1a} (Conditional independent $\Delta_k$): 
    \begin{equation}
    \underline{Y}_k^{g}\independent \Delta_{k} \mid (\overline{\mathbf{L}}_{k-1},\overline{A}_{k-1}=\overline{g}_{k-1}, \overline{Y}_{k-1}=1, \overline{\Delta}_{k-1}=0);
        \label{eq:condIC}
    \end{equation}
In Section \ref{sec:connection}, we introduce a related but distinct {conditional independent $\Delta_k$} assumption based on factual variables, which is closely connected to the classical survival assumption of conditional independent censoring.
    \item \textit{Assumption 2.2a} (Conditional treatment exchangeability): $$\underline{Y}_k^{g}\independent A_{k} \mid (\overline{\mathbf{L}}_{k-1},\overline{A}_{k-1}=\overline{g}_{k-1}, \overline{Y}_{k-1}=1,\overline{\Delta}_k=0);$$
    \item \textit{Assumption 2.3} (Consistency): $(\overline{\bL}_k^{g},\overline{\bC}_k^{g},\overline{Y}_k^{g})=(\overline{\bL}_k,\overline{\bC}_k,\overline Y_k) ~\text{if}~\bar{A}_{k-1}=\overline{g}_{k-1}$.
\end{itemize}

\begin{definition}[Model $\mathcal{M}_2$]
    Let $\mathcal{M}_2$ denote a model that satisfy right-censoring event irrelevance and Assumptions 1.4, 2.1a, 2.2a, 2.3.
\end{definition}

\begin{proposition}
     Under models $\mathcal{M}_2$, $\Psi_2^g$ is identified via equation \eqref{eq:ident}. Under models $\mathcal{M}_1\cup \mathcal{M}_2$, $\Psi_1^g$ is identified via equation \eqref{eq:ident}.
     \label{prop:model2}
\end{proposition}

{Conditional independent $\Delta_k$} implies that no event in $\bC_k$ can exert a direct effect on the outcome of interest (see a proof in Appendix D), including any effects mediated through $\underline{\bL}_{k}$. Consequently under model $\mathcal{M}_2$, both estimands are identified and yield the same value (i.e., $\Psi_1^g = \Psi_2^g$).
By contrast, when the full-data law $P$ lies in $\mathcal{M}_1$ but not in $\mathcal{M}_2$, we generally expect $\Psi_1^g \neq \Psi_2^g$ (see Section \ref{sec:identslip} for a further discussion).
Importantly, even when $P\notin \mathcal{M}_2$, $\Psi_2^g$ remains a well-defined target of inference, although it is no longer identified by \eqref{eq:ident}.
In particular, when {conditional independent $\Delta_k$} in $\mathcal{M}_2$ does \textit{not} hold, then $\Psi_2^g$ is not readily identified without making further assumptions. We discuss possible approaches below.

\subsection{Generalization and summary of estimand identification}
Thus far, we have considered two estimands $\Psi_1^g$ and $\Psi_2^g$ which are both based on potential outcomes. In Web Appendix A--D, we extend this to more general estimands, including those based on factual outcomes as well as estimands arising from randomized controlled trials. There we continue to use our classification of right-censoring events to derive clear assumptions and identifying formulas from the observed data. In Appendix F, we propose an extended definition of a right-censoring event to also capture events by time $k$ after which the investigator deems an individual's data unreliable or unusable, even if such events do not strictly prevent observation of future outcomes of interest. 

We have outlined sufficient conditions under which the various estimands considered in this work are all identified by the same formula given in \eqref{eq:ident} (see Table \ref{tab:identification} for a brief overview). In general, the choice of causal estimand should be guided by the underlying causal question. 
Even when some of the identification conditions are not satisfied, identification may still be achievable through the use of proxy variables (see \citealp{ying2022proximal,ying2024proximal} for a proposal in a non-causal setting).
In Web Appendix F, we present an approach that leverages proxy data to identify $\Psi_2^g$ when {conditional independent $\Delta_k$} is violated for special point treatment settings. Alternatively, one can conduct sensitivity analyses to quantify the extent to which conclusions might change under deviations from those assumptions to assess the robustness of the findings. We present one such analysis in Appendix E.

\begin{table}[h]
\small
    \centering
    \begin{tabular}{c|ccc}
        \hline
        \textit{Models} & \multicolumn{3}{c}{\textit{Estimand}} \\
        \hline
         $\mathcal{M}_1$ & $\Psi_1^g\coloneqq {E}(Y_k^{g,\bar{\bc}=0})$ & $\neq$ & $\Psi_2^g\coloneqq {E}(Y_k^{g})$ \\ 
         & Identified by \eqref{eq:ident} & & Not identified by \eqref{eq:ident} \\
        \hline
         $\mathcal{M}_2$ & $\Psi_1^g\coloneqq {E}(Y_k^{g,\bar{\bc}=0})$ & = & $\Psi_2^g\coloneqq {E}(Y_k^{g})$ \\ 
         & Identified by \eqref{eq:ident} & & Identified by \eqref{eq:ident}\\
        \hline
    \end{tabular}
    \caption{Identification results under different models defined herein.}
    \label{tab:identification}
\end{table}
\normalsize

\section{Connection to classical survival analysis}
\label{sec:connection}


Classical survival analysis literature typically do not distinguish between types of right-censoring events; instead, assumptions are usually stated in terms of being right-censored. As described in \cite{andersen2012statistical} (see also \citealp{KleinMoeschberger2003,klein2014handbook,tutz2016modeling}):
\begin{quote}
    ``The best known example of such \textit{incomplete observation} is \textit{right-censoring} in classical survival analysis; here, not all of a set of independent lifetimes are observed, so that for some of them it is only known that they are larger than some specific value.''
\end{quote}
This passage reflects a traditional survival-analysis view, which regards right-censoring as an observed data feature (the state of being right-censored) and formulates assumptions directly on that state. Our definition of right-censoring events provides insight into classical assumptions, namely, \textit{conditional independent censoring}.
In particular, a connection between this classical assumption and the set of assumptions we introduced above emerges when the events in $\bC_k$ are nuisance right-censoring events relative to a given estimand. 

Specifically, consider the estimand $\Psi_2^g$ whose identification we discussed in Section \ref{sec:identpsi2}. Rather than imposing the condition in \eqref{eq:condIC}, we can formulate the following version of conditional independence for $\Delta_k$ based solely on factual variables, which can also be used to identify $\Psi_2^g$. For each $k=1,\ldots,K$, consider the following assumption:
\begin{itemize}[label={},leftmargin=0mm]
    \item \textit{Assumption 2.1b.} (Conditional independent $\Delta_k$): 
\begin{equation}(\underline{Y}_k,\underline{\mathbf L}_k,\underline{A}_k)\independent \Delta_{k} \mid (\overline{\mathbf{L}}_{k-1},\overline{A}_{k-1}, \overline{Y}_{k-1}, \overline{\Delta}_{k-1}=0).\label{eq:condIC2}\end{equation} 
\end{itemize}
By induction and the monotonicity of $\Delta_k$, \eqref{eq:condIC2} is equivalent to $$(\underline{Y}_k,\underline{\mathbf L}_k,\underline{A}_k)\independent \bar\Delta_{k} \mid (\overline{\mathbf{L}}_{k-1},\overline{A}_{k-1}, \overline{Y}_{k-1}).$$ 
Note the difference between the \eqref{eq:condIC} and \eqref{eq:condIC2}: \eqref{eq:condIC2} involves only factual variables and thus connects directly to the missing-data and classical survival analysis literature, as we explain below. In contrast, \eqref{eq:condIC} is stated in terms of the potential outcomes; this formulation allows $\bC_k$ to affect $A_k$, whereas \eqref{eq:condIC2} does not permit that dependence. This distinction would not have been apparent if we had relied solely on the classical treatment of right-censoring, without engaging with the causal model.

 Nonetheless, when all events in $\bC_k$ are nuisance right-censoring events with respect to the estimand, some practitioners may appeal to the conditional independence of $\Delta_k$ as formulated in \eqref{eq:condIC2}. This assumption is analogous to the \textit{coarsening at random} assumption for monotone missing data in longitudinal studies. Specifically, if we temporarily treat $A_k\in \bL_k$, then \textit{conditional independent $\Delta_k$} 2.1b can be written as $(\underline{Y}_k,\underline{\mathbf L}_k)\independent \Delta_k\mid (\overline{\mathbf{L}}_{k-1},\overline{Y}_{k-1}=1,\overline{\Delta}_{k-1}=0)$, which is equivalent to the usual \textit{coarsening
at random} or \textit{conditional independent censoring} assumption for survival outcomes defined in Chapter 9.3 of \cite{Tsiatis2006} (see also \citealp{laan2003unified,Rotnitzky2014}).

\begin{remark}
Define $T$ as the failure time and $C$ as the observation-terminating time for a subject.  In classical survival literature, conditionally independent censoring states that $\lim_{\epsilon\rightarrow 0} P(C < t + \epsilon \mid C \geq t, T, \bar \bL(T)) = \lim_{\epsilon\rightarrow 0} P(C < t + \epsilon \mid C \geq t, T\geq t, \bar \bL(t)).$  Indeed if we let the time interval lengths shrink to zero, then $(\underline{Y}_k,\underline{\mathbf L}_k)\independent \Delta_k\mid (\overline{\mathbf{L}}_{k-1},\overline{Y}_{k-1}=1,\overline{\Delta}_{k-1}=0)$ would indeed be equivalent with this classical definition. 
\end{remark}

To this end, consider how this alternative condition can be used to achieve identification of $\Psi_2^g$. Specifically, consider a modified version of the conditional treatment exchangeability assumption posed previously in 2.2a:
\begin{itemize}[label={},leftmargin=0mm]
    \item \textit{Assumption 2.2b} (Conditional treatment exchangeability): $$\underline{Y}_k^{g}\independent A_{k} \mid (\overline{\mathbf{L}}_{k-1},\overline{A}_{k-1}=\overline{g}_{k-1}, \overline{Y}_{k-1}=1);$$
\end{itemize}
Assumption 2.2b corresponds to the standard conditional treatment exchangeability in the absence of any $\bC_k$ events. Framed this way, assumptions 2.1b and 2.2b separate the causal estimand from the handling of right-censored factual data, mirroring how missing data are often handled in causal inference literature \citep{ding2014identifiability,zhang2016causal,yang2019causal,kennedy2020efficient,levis2025robust,wen2025estimating}.
Based on this, we can define a model $\mathcal{M}_2'$ that encompasses Assumptions 1.4, 2.1b, 2.2b and 2.3. Proposition \ref{prop:model2} then continues to hold when $\mathcal{M}_2$ is replaced by $\mathcal{M}_2'$.

\section{Identity slippages}
\label{sec:identslip}
\sloppy
Estimation of the g-formula in \eqref{eq:ident} has been extensively studied under various treatment regimes. For time-to-event outcomes, \citet{Young2011} described the parametric g-formula, while \citet{wen2021parametric} described iterative conditional expectation estimators. Related work has investigated inverse probability weighted and doubly robust estimators \citep{robins2000correcting,hernan2000marginal,tran2019double,wen2022multiply,wen2023intervention}. Inverse probability weighted and doubly robust estimation has also been applied, for instance, to evaluate the effects of glucose-lowering strategies on the development or progression of albuminuria in patients with diabetes \citep{Neugebauer2014,sofrygin2019targeted}, and to assess grace-period treatment strategies in hypertension \citep{wanis2024grace}. We refer readers to these studies for detailed estimation methods based on \eqref{eq:ident}. A common feature in many of these works is that right-censoring events are intervened upon through their elimination; consequently, any resulting inferences pertain to the effect of this joint intervention. This motivates the following.

{Conditional independent $\Delta_k$} is violated when either (1) nuisance right-censoring events affects the outcome of interest, and/or (2) there are unmeasured common causes between those events and the outcome of interest. 
Here we show via simulation studies that in general $\Psi_1^g\coloneqq {E}(Y_k^{g,\bar{\bc}=0}) \neq \Psi^g_2 \coloneqq {E}(Y_k^{g})$ when {conditional independent $\Delta_k$} is violated (i.e., $P\notin \mathcal{M}_2$), even in absence of unmeasured common causes between right-censoring events and the outcome of interest ($P\in \mathcal{M}_1$). Consequently, this discrepancy can lead to identity slippage \citep{sarvet2023interpretational}, where researchers may inadvertently draw conclusions about the average treatment effect of an intervention on an outcome based on estimates for $\Psi_1^g$ when $\Psi_2^g$ was the initial target parameter of interest and $\mathcal{M}_1$ was assumed in the analysis. 

\sloppy
For simplicity, we consider a point treatment process where $A_k=\emptyset$ for $k>0$ and a scalar observation-terminating event at each time $k$, $C_k$. 
In this example, the only right-censoring event is $C_k$, and the treatment regime of interest is $g=a_0$ with $a_0=1$.
Consider the following data generating mechanism: $(L_{10},L_{20}, A_0, C_{1}, Y_1,L_{11},L_{21}, C_{2}, Y_2, \ldots, C_{5},Y_5)$
where $L_{1k}$ and $L_{2k}$ are time-dependent confounders at time $k$. At baseline, we draw $L_{10}\sim \text{Ber}(0.5)$, $L_{20}\sim \text{Ber}(0.5)$ and assign treatment according to $A_0 \sim \text{Ber}\{\expit(1-2L_{10}+L_{20}+L_{10}L_{20})\}$. 
Moreover, $C_k$ and $Y_k$ at each time $k$ ($k\geq 1$) are simulated from:
\begin{align*}
C_{k} \sim \text{Ber}\{\expit(\textcolor{purple}{\gamma}-A_0+L_{1,k-1}+L_{2,k-1})\}~ \text{if}~C_{k-1}=0;~~~
    & C_{k} = 1~ \text{if}~C_{k-1}=1\\
    Y_k \sim \text{Ber}\{m({A_0},\bar{\mathbf{L}}_k, \bar{C}_k)\}~\text{if}~Y_{k-1}=1;~~~ &
    Y_k = 0~\text{if}~Y_{k-1}=0
\end{align*}
where $m(A_0,\bar{\mathbf{L}}_k, \bar{C}_{k})=\expit(-1+2A_0+2L_{1,k-1}-2L_{2,k-1}+2L_{1,k-1}L_{2,k-1}+\textcolor{purple}{\alpha} C_{k})$. 
When $Y_{k}=1$, time-varying confounders at time $k$ ($k\geq 1$) are simulated from: $L_{1k} \sim  \text{Ber}\{\expit(A_0+L_{1,k-1}-L_{2,k-1} )\}$, and $L_{2k} \sim  \text{Ber}\{\expit(-2-2A_0-L_{1k})\}$ if $L_{2,k-1}=0$, and is deterministically set to $1$ if $L_{2,k-1}=1$.
For the estimand $\Psi_1^g$, $C_k$ is a non-nuisance right-censoring event, whereas for $\Psi_2^g$, $C_k$ is a nuisance right-censoring event.

In the data generating mechanism, $\gamma$ controls the amount of censoring and $\alpha$ quantifies the extent to which censoring affects the conditional probability of survival. A negative value of $\alpha$ suggests that experiencing the right-censoring event reduces the probability of survival. We consider a range of values for $\alpha$ and $\gamma$ in simulated datasets of sample size $N=10^7$.

\begin{figure}[htbp]
    \centering
    \begin{minipage}{\linewidth}
        \centering
        \fbox{\includegraphics[width=1\linewidth]{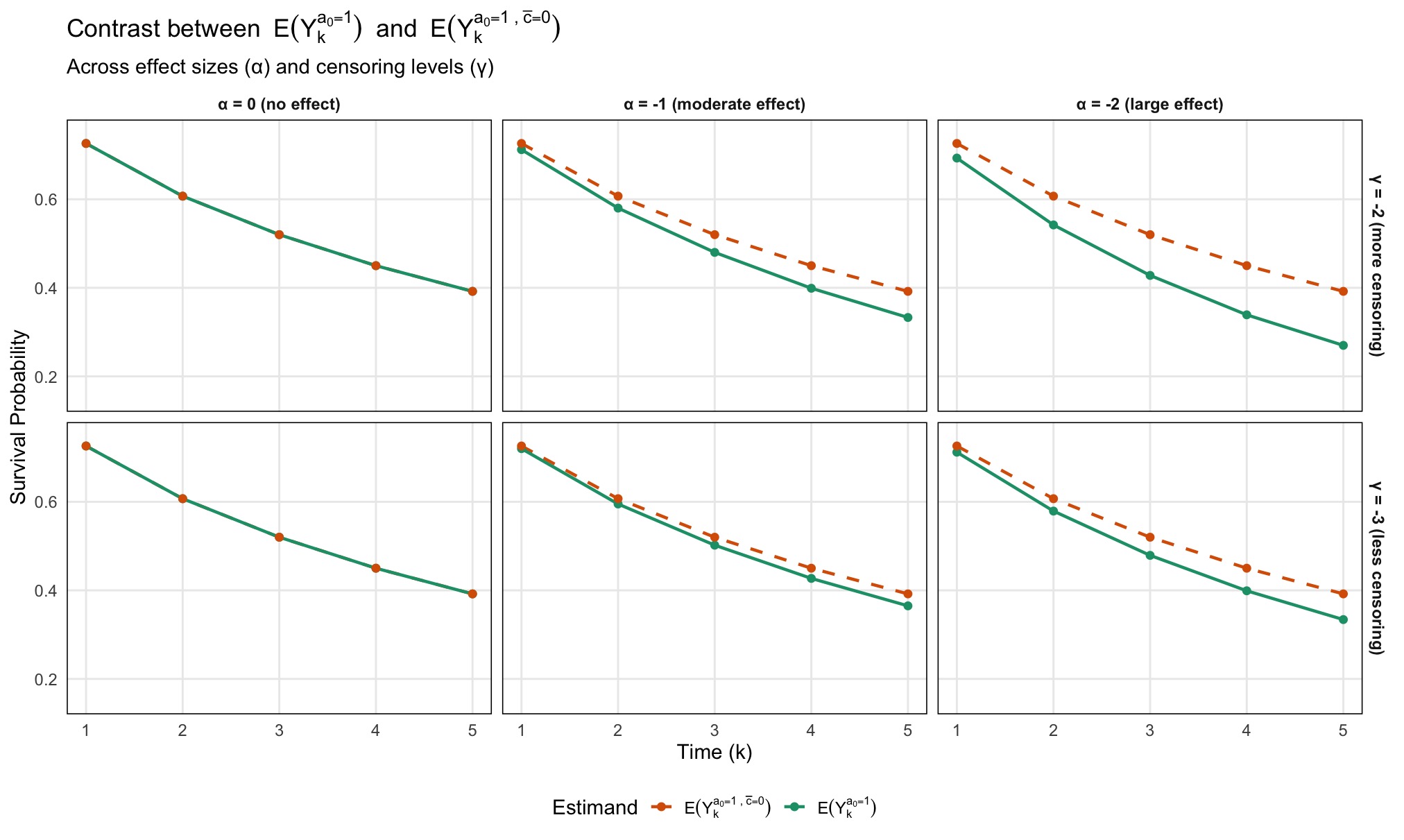}}
        \caption{Survival probabilities of ${E}(Y_k^{a_0=1})$ and ${E}(Y_k^{a_0=1,\bar{\bc}=0})$ under different conditions of $\alpha$ and $\gamma$ across time points for sample size $N=10^7$.}
        \label{fig:survival_alpha_gamma}
    \end{minipage}  \\ \bigskip
    \begin{minipage}{\linewidth}
        \centering
        \fbox{\includegraphics[width=0.65\linewidth]{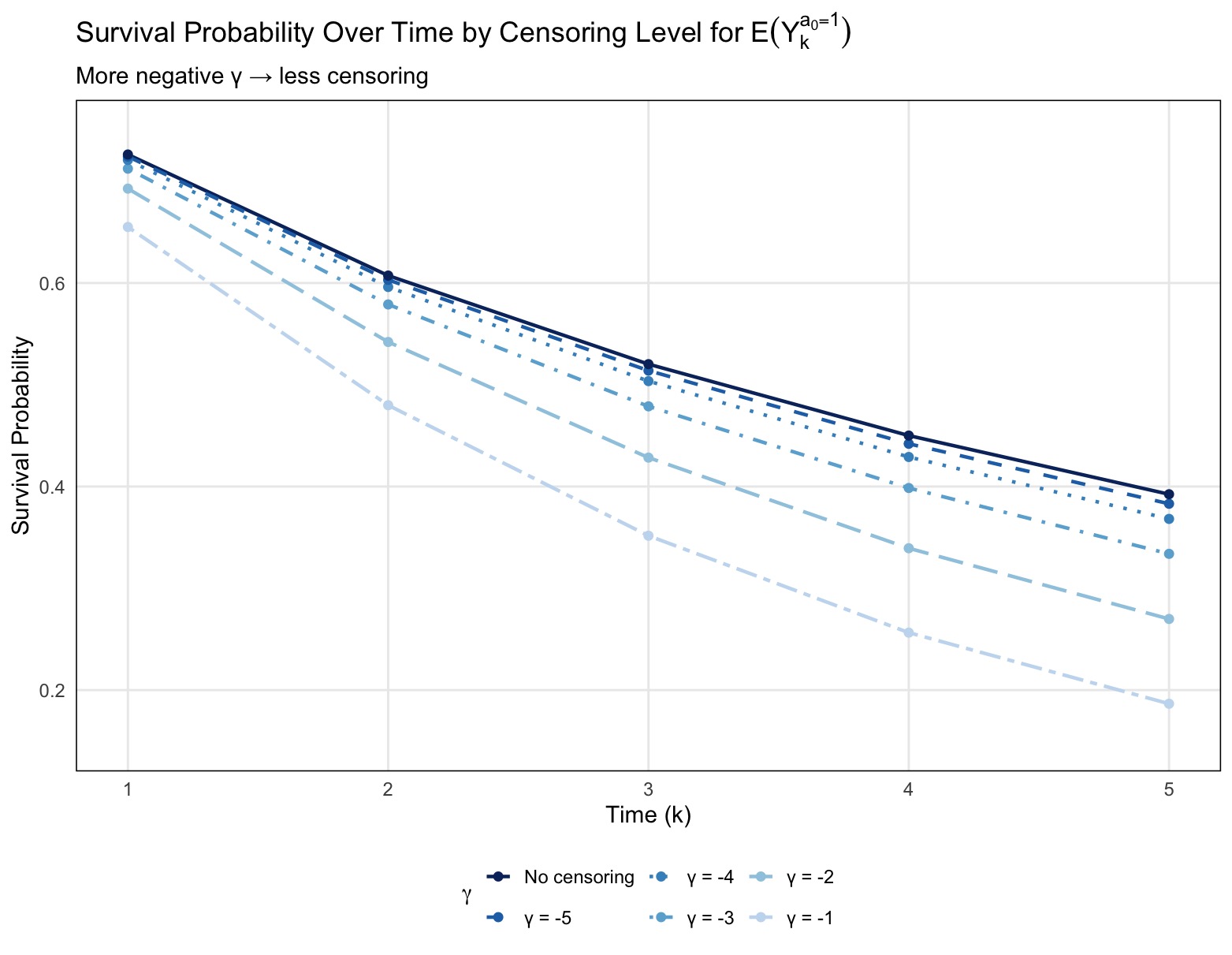}}
        \caption{Survival probabilities of ${E}(Y_k^{a_0=1})$ under different amounts of censoring across time points for sample size $N=10^7$ when $\alpha=-2$. \textit{In ``No censoring'' setting, ${E}(Y_k^{a_0=1}) = {E}(Y_k^{a_0=1,\bar{\bc}=0}).$}}
        \label{fig:survival_gamma}
    \end{minipage}
\end{figure}

The results are presented in Figures \ref{fig:survival_alpha_gamma} and \ref{fig:survival_gamma}.  Under {conditionally independent $\Delta_k$} (when $\alpha=0$),  theory predicts that that $\Psi_1^g=\Psi_2^g$, which is confirmed by our results in Figure \ref{fig:survival_alpha_gamma}. However, as the dependence of survival probability on censoring status increases, the two estimands begin to diverge. 
Furthermore, when $\alpha\neq 0$, this divergence is more pronounced under a greater amount of right-censoring. For instance, when $\gamma=-2$ (approximately 56\% observations are right-censored by time 5), the divergence is greater at all time points compared to when $\gamma=-3$ (approximately 28\% are right-censored by time 5).
Conversely, in the absence of right-censoring event the estimands coincide ($\Psi_1^g=\Psi_2^g$).
Figure \ref{fig:survival_gamma} highlights this trend for a fixed strong effect $\alpha$ ($\alpha=-2$): greater amount of right-censoring produces a larger gap between $\Psi_1^g$ and $\Psi_2^g$.

Whereas $\Psi_1^g$ remains identifiable from observed data via the g-formula, $\Psi_2^g$ is not readily identifiable without making additional assumptions.
To illustrate identifiability, we report results from a parametric g-formula estimator based on \eqref{eq:ident} (see Web Appendix F for inverse probability weighted estimation results); more robust estimators (e.g., doubly robust estimators) may be used in practice. 
With a sample size of $N=1000$ across 1000 simulated datasets, the parametric g-formula estimator produced nearly unbiased estimates of $\Psi_1^g\coloneqq E(Y_k^{a_0=1,\bar{\bc}=0}) = (0.7263, 0.6073, 0.5203,0.4501, 0.3924)$ for $k=1,\ldots,5$ under correct specification of all nuisance models. 
Specifically, the estimator yielded values of $(0.7268, 0.6077, 0.5207, 0.4506, 0.3930)$ with corresponding standard errors $(0.0125,~0.0152,~0.0175,~0.0192,~0.0204)$.
In Appendix E, we present a sensitivity analysis for $\Psi_2^g\coloneqq E(Y_k^{a_0=1})$. We also use this simulation study to examine the performance of the method described in Web Appendix F for estimating $\Psi_2^g$ under violations of {conditional independent $\Delta_k$}; the results indicate that the proposed estimator performs well and is nearly unbiased in these settings.

Our simulation shows that $\Psi^g_1$ coincides with $\Psi^g_2$ when $\bC_k=\emptyset$ or when $\bC_k$ does not have a direct effect on $Y_k$. In our simulation setting, $C_k$ is the only right-censoring event, and when this right-censoring event is associated with a lower probability of survival, we observe that $\Psi^g_1\geq \Psi^g_2$. In other words, the probability of survival under an intervention that sets $A_0$ to $a_0=1$ and eliminates the right-censoring event is at least as large as the probability of survival under an intervention that sets $A_0$ to $a_0=1$ without intervening on censoring. 
Web Appendix E presents results for causal contrasts, which align with the findings reported here.

\section{Concluding remarks}
The choice of estimand should be guided by underlying scientific objectives, as it determines how the right-censoring events are classified and consequently, how they should be handled.
In this paper, we proposed a general definition of right-censoring events, and provided a classification that separates them into nuisance and non-nuisance categories. 
A nuisance right-censoring event is one that the investigator's causal question and estimand does not aim to eliminate or prevent. In contrast, a non-nuisance right-censoring event is integral to the definition of the estimand.

Recent advances in causal inference propose that observation-terminating events $\bC_k$ need not be framed as a missing data problem, but rather as part of an intervention in which all right-censoring events are eliminated.
We argue instead that whether to treat such observation-terminating events as interventions depends on the study objective. 
For instance, identification of the estimand given by ${E}(Y_k^{g,\bar{\bc}=0})$ requires intervening on all observation-terminating events (considered non-nuisance right-censoring here) and adjusting for all common causes of these events and the outcome. In contrast, identification of ${E}(Y_k^{g})$ does not require intervening on observation-terminating events, but still requires adjustment for all common causes of nuisance right-censoring events and the outcome.

The subtlety in handling events in $\bC_k$ is that, whereas estimands such as intention-to-treat and per-protocol typically rest on different identifying assumptions and therefore different identification formulas, estimands defined in the presence of observation-terminating events can, under different sets of assumptions and models, have identical identifying formulas. 
However, as noted above and emphasized again here, identifying ${E}(Y_k^{g})$ relies on a different set of assumptions than those required for ${E}(Y_k^{g,\bar{\bc}=0})$, and consequently, these two quantities may differ in practice.
Ultimately, handling right-censoring events should follow from the estimand of interest rather than convenience or ease of identification.
If a right-censoring event is not directly intervenable (see \citealp{young2020causal}), consistency might not hold due to the ambiguity in counterfactuals arising from the multiple pathways or mechanisms involved. 
If this was indeed the case for observation-terminating events, then defining estimands based on the potential outcome $Y_k^{g,\bar{\bc}=0}$ might be questionable.

Our findings highlight the importance of understanding right-censoring events in each study. 
Such knowledge is indispensable for accurately accounting for the shared common causes between the censoring mechanisms and the outcome of interest. 
Without this understanding, the validity of the estimand and the resulting inferences may be compromised. Thus, careful consideration of censoring mechanisms and their relationships with the outcome is essential for robust causal inference in the presence of right-censored data.

\bibliographystyle{biorefs}
\bibliography{refs}


\end{document}